\documentclass[twoside]{article}
\usepackage{fleqn,espcrc2,epsfig}

\newcommand{\AmS}{{\protect\the\textfont2
  A\kern-.1667em\lower.5ex\hbox{M}\kern-.125emS}}

\title{Composite reweighting SU(2) QCD at Finite Temperature}

\author{ P. R. Crompton
\address{Dept. of Physics and Astronomy, University of Glasgow, G12 8QQ, 
Scotland, UK.}}

\begin{document}
\pagestyle{empty}
\setcounter{topnumber}{1}

\begin{abstract}
{The Glasgow reweighting method is evaluated for SU(2) lattice gauge theory 
at nonzero $\mu$ and finite $T$. We 
establish that the 'overlap problem' of SU(3) measurements, in which the 
transition points determined from thermodynamic observables have an 
unphysical dependence on the value of $\mu$ used to generate ensembles 
for reweighting, persists for SU(2). By 
combining the information from different lattice ensembles we alleviate 
sampling bias in the fugacity expansion, and identify the Lee Yang zeros 
associated with the transition to a high density phase that can plausibly be 
associated with diquark condensation. 
We also confirm the existence of a line of first order 
transitions above a critical 
point in the $T-\mu$ phase plane previously predicted by effective chiral 
lagrangian calculations.}
\end{abstract}

\maketitle

\section{Introduction}
Recent speculation over BCS type-instabilities in the Fermi surface at high 
density \cite{1}\cite{2} has led to resurgence of interest in two colour QCD.
Lattice models with pseudoreal representations are attractive candidates for 
simulation at finite density since for 
such theories the Dirac matrix is positive definite, which permits the use of 
existing Monte Carlo techniques at $\mu \neq 0$. For 
SU(3) at finite density the lattice action is otherwise complex 
\cite{3}, and so the 
probabilistic importance sampling step of such methods is therefore undefined. 
Reweighting methods have proved a useful means of addressing this issue, 
where the $\mu$ dependence of the grand canonical partition function $Z(\mu)$ 
can be made semi-analytic in a fugacity expansion, as with the 
Glasgow method \cite{4}. 
The complex action issue is thus avoided, by generating a 
lattice ensemble in an accessible regime of the parameter space 
(eg. $\mu =0$ for SU(3)). Naively, one 
might anticipate that the specific lattice ensemble used in the reweighting 
has little impact on numerical evaluations of the expansion. 
In fact quite the reverse is true. For the SU(3) model with dynamical quarks, 
even at intermediate coupling where one might expect
 thermal fluctuations to enhance the frequency of sampling physically
 relevant states, an ensemble generated at $\mu=0$ reproduces a similar phase 
structure to the quenched model \cite{5}\cite{6}. 
This has severe consequences since quenched 
finite density QCD is understood to be the zero flavor limit 
($n \rightarrow 0$) of a theory with equal numbers of quarks and conjugate 
quarks. The lowest lying baryonic state in the model is thus the 
unphysical ``baryonic pion'', 
formed from quark-conjugate quark pairs rather than the 
lightest three quark state \cite{7}\cite{8}. However, this is of no 
consequence in two colour QCD \cite{9} as the baryonic pion and baryon 
propagators are equivalent at $\mu \neq 0$.

Pseudoreal models are not entirely free from the effects of the reweighting 
overlap pathologies, however. Models with quarks in the adjoint 
representation (with $n$ odd) suffer from the related reweighting pathology: 
the sign problem. Since ${\rm{det}}M(\mu)$ is always real for two colour QCD, 
although reweighting is no longer mandatory it provides a 
useful opportunity
 to investigate the overlap issue (where the correct physics may be more 
easily extracted from lattice measurements by conventional means), and to 
quantify the signatures of the sampling numerical discrepancies therein.
\vspace{-0.1in} 
\section{Symmetries of the SU(2) Lattice Action}
The $n-$flavor symmetry of SU(2) QCD given by quarks in their fundamental 
representation $\bf{2}$, is $SU(2n)$ rather than 
$U(n)_{L} \times U(n)_{R}$ as might be 
anticipated, as is demonstrated by a change of basis in the free 
field langrangian.
\begin{equation}
\mathcal{L} \,= \,\overline{\psi} \gamma_{\nu} D_{\nu} \psi \, = \, 
i \Psi^{\dagger} \sigma_{\nu} D_{\nu} \Psi 
\end{equation}
\begin{equation}
\Psi = \left( \begin{array}{c} 
 \psi_{L} \\
 \sigma_{2}\tau_{2}\psi_{R}^{*} \end{array}\right)
\end{equation}
where $\psi$ is a $\bf{2}$ doublet. The inclusion of explicit symmetry 
breaking terms in $m$ and $\mu$ can be shown similarly to lead to the symmetry 
breaking patterns tabulated above \cite{9}\cite{10}. 
\begin{eqnarray}
\overline{\psi}\psi & = & \frac{1}{2} \Psi^{T} \sigma_{2} \tau_{2} 
\left( \begin{array}{cc} 
 0 & -1 \\
 1 & 0 \end{array}\right)\Psi + \,h. \,c. \label{pbp} \\
\overline{\psi} \gamma_{o}\psi & = & \Psi^{\dagger} \left( \begin{array}{cc} 
 1 & 0 \\
 0 & -1 \end{array}\right)
\Psi \label{pp}
\end{eqnarray}

\begin{table}
\begin{center}
\begin{tabular}{|c|c|c||} \hline
 & $m, \mu =0$ & $m \neq 0,\mu =0 $ 
\\ \cline{1-1} \cline{2-2} \hline 
Cont. & $SU(2n)$ & $Sp(2n)$ \\ \hline
Latt. & $U(2n)$ & $O(2n)$ \\ \hline
 & $m = 0, \mu \neq 0$ & $m,\mu \neq 0$ 
\\ \cline{1-1} \cline{2-2} \hline 
Cont. &  $SU(n)_{L} \times SU(n)_{R}$ & 
$SU(n)_{V}$ \\ \hline
Latt. & $U(n)_{V} \times U(n)_{A}$ & 
$U(n)_{V}$ \\ \hline
\end{tabular}	
\end{center}
\caption{Global $n-$flavor symmetries of the two colour QCD action, 
in the continuum and with staggered quarks for $m, \mu \neq 0$.}
\end{table}

Naturally, the lattice model (with Kogut-Susskind fermions), follows a 
somewhat similar symmetry breaking scheme, having a manifest 
global $U(2n)$ symmetry. In the continuum limit, for the choice of $n=1$, 
this lattice action corresponds to 8 physical flavors through the well-known 
doubling of fermionic modes \cite{11}.

At $m, \mu \neq 0$ it is argued in \cite{9} that since the number of the 
Goldstone modes differs for nonzero expectations of $\psi\psi$ and 
$\overline{\psi}\psi$, that a phase transition occurs at 
$\mu \geq \frac{1}{2}m_{\pi}$ corresponding to the point at which 
the number density of quarks becomes nonzero and $U(n)_{V}$ is spontaneously 
broken. 
It is then further argued with the Landau free energy 
in \cite{12}, that since the number of Goldstone modes at high density is 
odd, that the transition to the 
free quark phase at finite temperature is necessarily first order.

\section{Glasgow Method}
For the Glasgow reweighting method, the $\mu$ dependence of the lattice 
action is made analytic through the formulation of the fugacity expansion, 
where $z \equiv {\rm{exp}}(\mu/T)$. 
This constitutes the characteristic fugacity 
polynomial which is formed from the propagator matrix $P$, defined through the 
fermion matrix $M$ \cite{13}. Where $P$ is 
written in terms of the matrices which contain links between 
lattice sites in the spatial directions $G$, 
and forward and backward in the time direction $V$ and $V^{\dagger}$,
\begin{equation}
2iM = 2im + G + V e^{\mu} + V^{\dagger}e^{-\mu} 
\end{equation}
\begin{equation}
P = \left( \begin{array}{cc} 
 -( G + 2im ) & 1  \\
-1	       & 0   \end{array}\right) V
\label{P}
\end{equation}
\begin{eqnarray}
{\rm{det}} M  & = & {\rm{det}}  ( G + 2im + 
V^{\dagger}e^{-\mu} + Ve^{\mu} ) \\
& = & e^{n_{c}n_{s}^{3}n_{t}\mu} \,\,{\rm{det}}  ( P - e^{-\mu} ) \nonumber \\
& = & e^{n_{c}n_{s}^{3}n_{t}\mu}\sum_{n=0}^{2n_{c}n_{s}^{3}n_{t}}
c_{n}e^{-n\mu} 
\end{eqnarray}
with $n_{s}^{3}n_{t}$ the lattice volume and $n_{c}$ the number 
of colours in the expansion. 
Since $V$ is an overall factor of $P$ the order of the expansion is reduced 
by exploited the unitary symmetry $Z_{n_{t}}$ defined by multiplying the 
timelinks in $V$ by $e^{2\pi ij / n_{t}}$, where $j$ is an integer. 
Since $n_{t} = 1/T$, the Grand Canonical Partition function $Z(\mu)$ is thus 
given defined as an expansion in terms of the fugacity variable and 
the canonical partition functions $Z_{n}$.
\begin{eqnarray}
Z(\mu) & = & \int DU \,\, {\rm{det}} M(\mu) \,\, e^{-S_{g}} \\ 
& = & \sum_{n} \,\, Z_{n} \,\, e^{n\mu/T}
\end{eqnarray}
By reweighting this expansion an arbitrary normalisation to $Z(\mu)$ 
is introduced, though this leaves the 
analytic determination of thermodynamic variables unaffected. 
\begin{eqnarray}
\frac{Z(\mu)}{Z(\mu_{o})} & = & \frac {\int DU \,\, 
{\displaystyle{\frac{ {\rm{det}} 
M(\mu) }{ {\rm{det}} 
M(\mu_{o}) }}}
\,\,\, {\rm{det}} M(\mu_{o}) \,\, e^{-S_{g}}} {\int DU 
\,\, {\rm{det}} M(\mu_{o}) \,\, e^{-S_{g}}} 
\nonumber \\ 
\label{norm}
 & = & \left\langle {\frac{{\rm{det}} M(\mu)}{{\rm{det}} M(\mu_{o})}} 
\right\rangle_{\mu_{o}} \\
\frac {Z_{n}} {Z(\mu_{o})} & = & \frac {\int DU {\displaystyle{\frac {c_{n}} { {\rm{det}} 
M(\mu_{o}) }}}
\,\, {\rm{det}} M(\mu_{o}) \,\, e^{-S_{g}}} {\int DU 
\,\, {\rm{det}} M(\mu_{o}) \,\, e^{-S_{g}}} \\
\label{aver}
& = & \left\langle {\frac{{c_{n}}}{{\rm{det}} M(\mu_{o})}}
 \right\rangle_{\mu_{o}} \label{ens} \nonumber
\end{eqnarray}
However, the reliability of the ensemble-averaging is 
strongly affected when the ratio of the ratio in eqn.(\ref{aver})
differs greatly from one. The canonical partition functions 
can, in general, only be reliably 
determined for a small series of terms in 
the expansion, centered on the term of order $n(\mu_{o})$. 
This effect and the reliability of the averaging elsewhere can be  
established for two-colour QCD 
by measuring the ratio of the ensemble-averaged expansion 
coefficients between two or more ensembles generated at different 
values of $\mu_{o}$. Having then 
identified  $n(\mu_{o})$ for several ensembles, 
our composite reweighting method then consists of 
rescaling the expansion coefficients from different 
ensembles through these 
ratios (where the ensemble-averaging is effective).  
The bias introduced through reweighting the expansion 
can thus be systematically 
alleviated, and thermodynamic 
observables more reliably determined \cite{14}. 
\subsection{Thermodynamic Observables}
The eigenvalues $\lambda_{n}$ 
of $P$ naturally share the symmetries of $V$, most notably 
$\lambda \rightarrow 1/\lambda^{*}$ relating $P$ to $P^{-1}$ up to 
a unitary transformation. 
Since SU(2) with quarks in the fundamental representation is pseudoreal 
it can also be shown that $\lambda \rightarrow \lambda^{*}$. By rewriting the 
expansion in the variable $y = z + 1/z$ the order can be further reduced by a 
factor of two to reduce rounding errors in the numerical implementation 
\cite{15}. 
The quark number density $n$ and its associated susceptibility $\chi_{n}$ 
for this expansion is then both easily evaluated from the 
expansion, and in addition is also 
readily amenable to composite reweighting approach described above. 
\begin{eqnarray}\langle \, n \, \rangle & = &  
\frac{T}{n_{s}^{3}} \frac{\partial \,{\rm{ln}} Z(\mu)}{\partial \mu}  \\
& = & 
\frac{\sum_{n = 0}^{n_{c}n_{s}^{3}} 
n \,\, {\rm{sinh}} {\displaystyle (-\frac{\epsilon_{n}-n\mu}{T})}}
{\sum_{n = 0}^{n_{c}n_{s}^{3}} 
{\rm{sinh}} {\displaystyle (-\frac{\epsilon_{n}-n\mu}{T})}} \\
\label{arse}
\langle \, \chi_{n} (\mu) \, \rangle & = &  \langle \,n^{2}\,\rangle - \langle \,n\, \rangle^{2} 
\end{eqnarray}
Similarly the zeros $\alpha_{n}$ 
of $Z(\mu)$ are readily identified from the expansion, 
which as Lee and Yang showed with an Ising ferromagnetic system, correspond 
to a phase transition in the thermodynamic limit 
wherever a zero approaches the real axis in the complex-$z$ plane \cite{16}. 
\begin{equation}
Z(\mu) \,\,\,\propto \,\,\,e^{-n_{c}n_{s}^{3}n_{t}\mu} \,\,
\prod_{n=1}^{n_{c}n_{s}^{3}} \,\,( e^{n_{t} \mu} - \alpha_{n} ) 
\end{equation}
\section{Results}
\subsection{Intermediate Coupling}
We generated a total of seven ensembles at consecutive values of 
$\mu_{o}$ ranging from $\mu_{o} = 0.3 - 1.1$, at $\beta_{c} = 1.5$ both for 
$4^{4}$ and $6^{3}4$ lattice volumes. 
From these lattice ensembles we evaluated the Lee-Yang zeros, 
quark number density susceptibility $\langle \chi_{n} \rangle$, and 
$\langle \overline{\psi} \psi \rangle$ using a conventional stochastic 
approach. 

For both our 
measurements at $\beta =1.5$ and $\beta =2.3$, 
$\langle \overline{\psi} \psi \rangle$ decreases 
gradually to zero over the range of values of $\mu$ we generated for 
$m = 0.05$. However, 
since $\langle \overline{\psi} \psi \,(\mu\!=\!0) \rangle$ is considerably 
smaller in the latter measurement, plausibly $U(1)_{A}$ is spontaneously 
broken in the chiral limit in the former case for the volumes we used.
It then follows that there should be a corresponding transition in the 
$m - \mu$ plane at $\mu_{c} \sim \frac{1}{2}m_{\pi}$ (as we argued in Sec 2), 
which we were able to identify from our Lee-Yang zeros measurements using 
composite reweighting. An 
unphysical $\mu_{o}$ dependence dominates our measurements prior to composite 
reweighting at $\beta =1.5$ and 
is tabulated in Table 2, along with the convergence of our 
measurements after composite reweighting 
as we increase the number of included ensembles.

For an ensemble generated at $\mu_{o} = \mu_{c}$ we believe the sampling 
should be effective enough to circumvent the need for 
composite reweighting. There is 
evidence to support this with the ensemble we generated at 
$\mu_{o}=0.3 \sim \mu_{c}$ 
in Fig.1, which shows more evidence of a transition 
(where the zeros consistently approach the real axis) at 
$\mu_{c} \sim \frac{1}{2}m_{\pi}$ 
than the other ensembles we generated at $\beta =1.5$. However, since we are 
unable to accurately quantify which values of $\mu_{c}(\mu_{o})$ are the more 
valid from our jacknife 
error estimates of the Lee-Yang zeros, and the unphysical 
$\mu_{o}$ dependence of our measurements persists for ensembles 
generated at values of $\mu_{o}$ arbitrarily close to $\mu_{c}$, 
we found it is more effective to sample the expansion 
coefficients accurately by 
generating a covering series of ensembles with our composite reweighting 
method.
 
In varying the lattice volume $V$ 
and $\beta$, we can confirm that this unphysical 
$\mu_{o}$ dependence in our measurements behaves as we would expect 
of a reweighting overlap(sign) problem. The expectation of the 
sign of the Monte Carlo measure (which is treated as an observable for 
reweighting in the Potts model), 
shows a $\beta$ and $V$ dependence of the form,
\begin{equation}
\langle sgn \rangle = \frac{Z}{Z_{||}} = {\rm{exp}}(-\beta V \Delta f)
\end{equation}       
where $Z_{||}$ is the partition function of the ensemble modified to exclude 
the sign problem amenable to a Monte Carlo approach, and $\Delta f$ 
the difference in free 
energy densities between ensembles \cite{17}. This effect is seen Tables 2 
and 3 where the imaginary part of the zeros nearest the real axis 
evaluated from the ensemble at generated at $\mu_{o} \sim \mu_{c}$ is 
comparatively smaller (and therefore more convincing) as $V$ is increased. 
Similarly, this Lee-Yang zeros effect becomes more pronounced as we increase 
$\beta$.

We are able to determine the range of values of $\mu_{o}$ over which to 
generate ensembles for our effective sampling strategy, from 
the jacknife error estimates for the ensemble-averaged expansion 
coefficients, which give us $n(\mu_{o})$. The 
largest of the coefficients $c_{2n_{c}n_{s}^{3}}$  
(related to the canonical 
partition function for the filled lattice) 
is of order one for $\mu_{o}=1.2$, and the lattice therefore saturated. 
Our quark number density susceptibility measurements $\langle \chi_{n} (\mu) 
\rangle$ become singular at $\mu_{c}$ 
as we include more ensembles in the composite 
reweighting across this range, indicating that the transition at $\mu_{c} \sim 
\frac{1}{2} m_{\pi}$ for $\beta =1.5$ is first order, Fig 4. 
We are also able to identified a 
second smaller peak in these measurements which 
corresponds to the point at which 
the expectation of the diquark falls off in existing condensate 
measurements. As saturation is 
approached at $\mu = 1.2$ the diquark condensate thus evaporates in a 
less well determined transition driven by Fermi statistics \cite{18}\cite{19}.

\begin{figure}
\centering{\epsfig{file=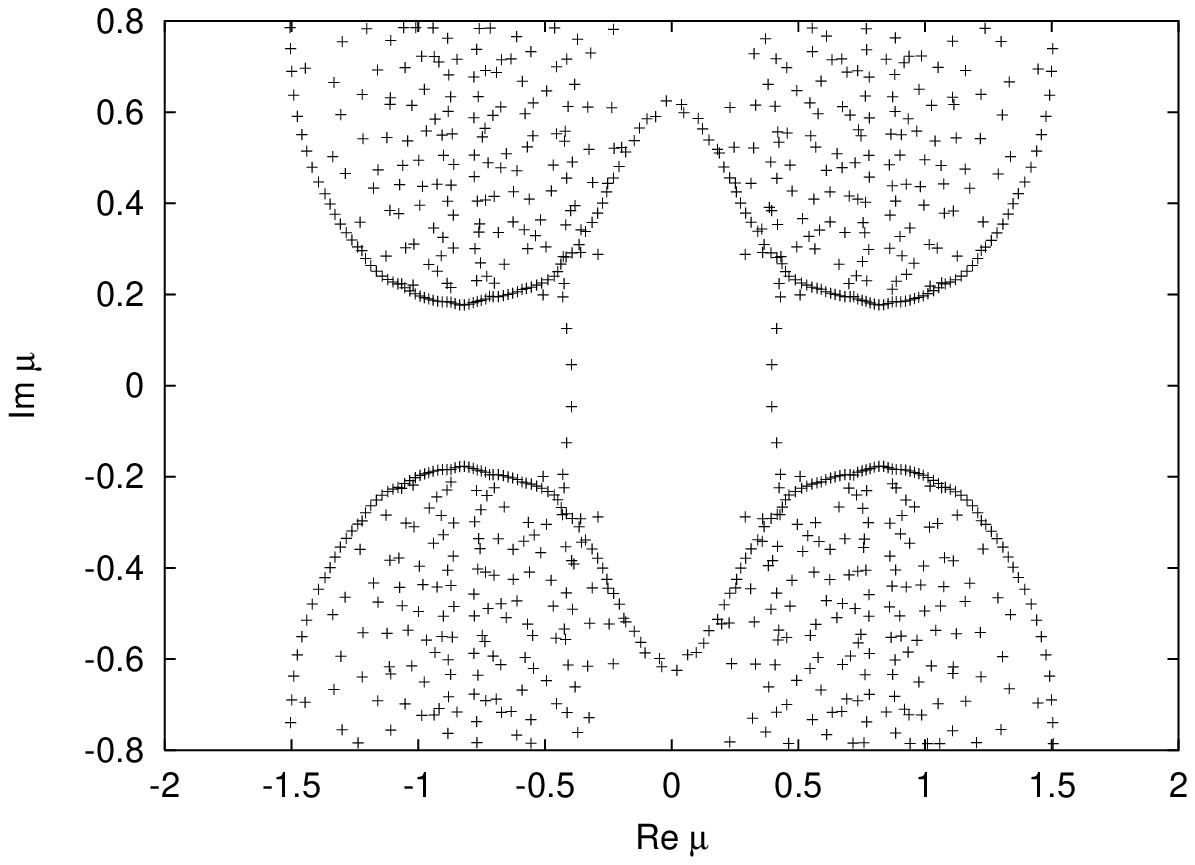, height=0.28\textheight}}
\vspace{-1.5cm}
\caption{Lee Yang zeros evaluated in the complex $\mu$ plane 
($\eta_{n} = T \, ln \, \alpha_{n} $) for a $6^{3}4$ lattice 
at $\beta=1.5$ from an ensemble generated at $\mu_{o}$ = 0.3.}
\label{fig:2}
\centering{\epsfig{file=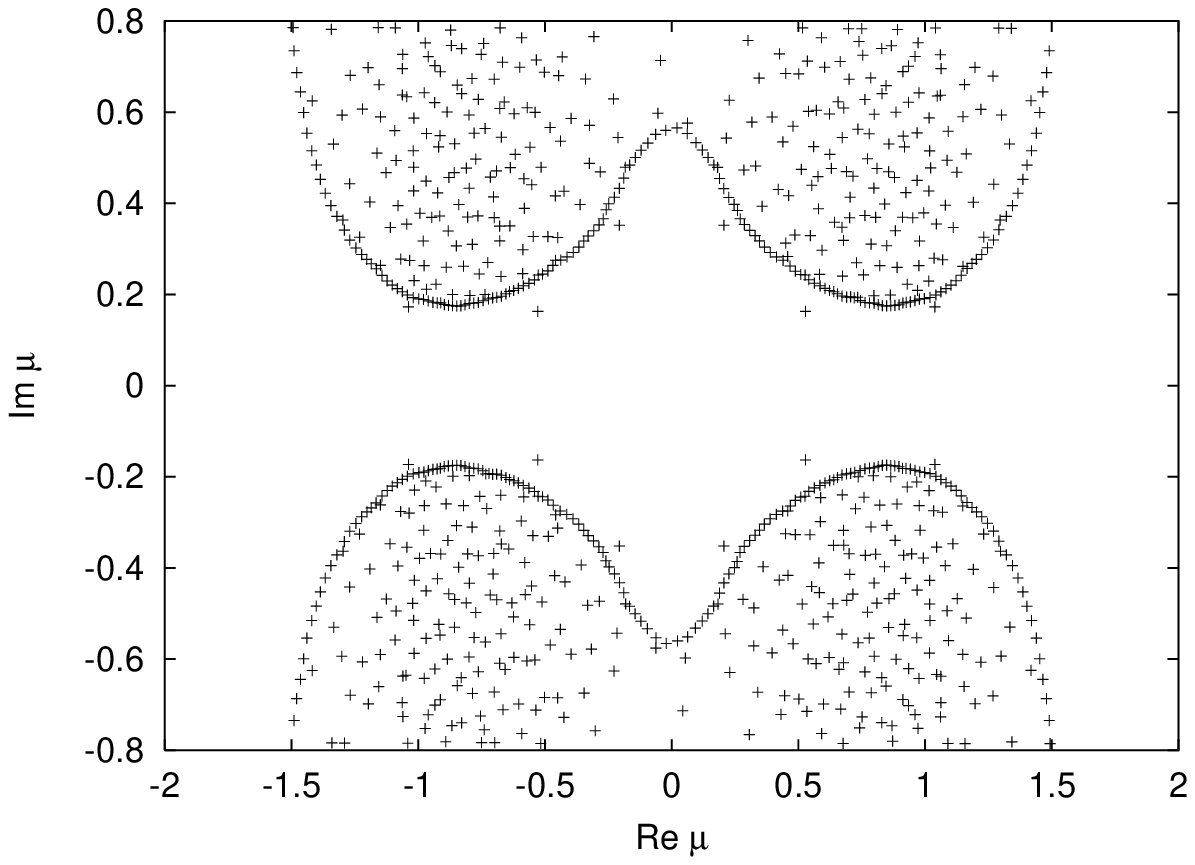, height=0.28\textheight}}
\vspace{-1.5cm}
\caption{Lee Yang zeros evaluated in the complex $\mu$ plane for a 
$6^{3}4$ lattice at $\beta=1.5$ from an ensemble generated at $\mu_{o}$=0.5.}
\label{fig:3}
\centering{\epsfig{file=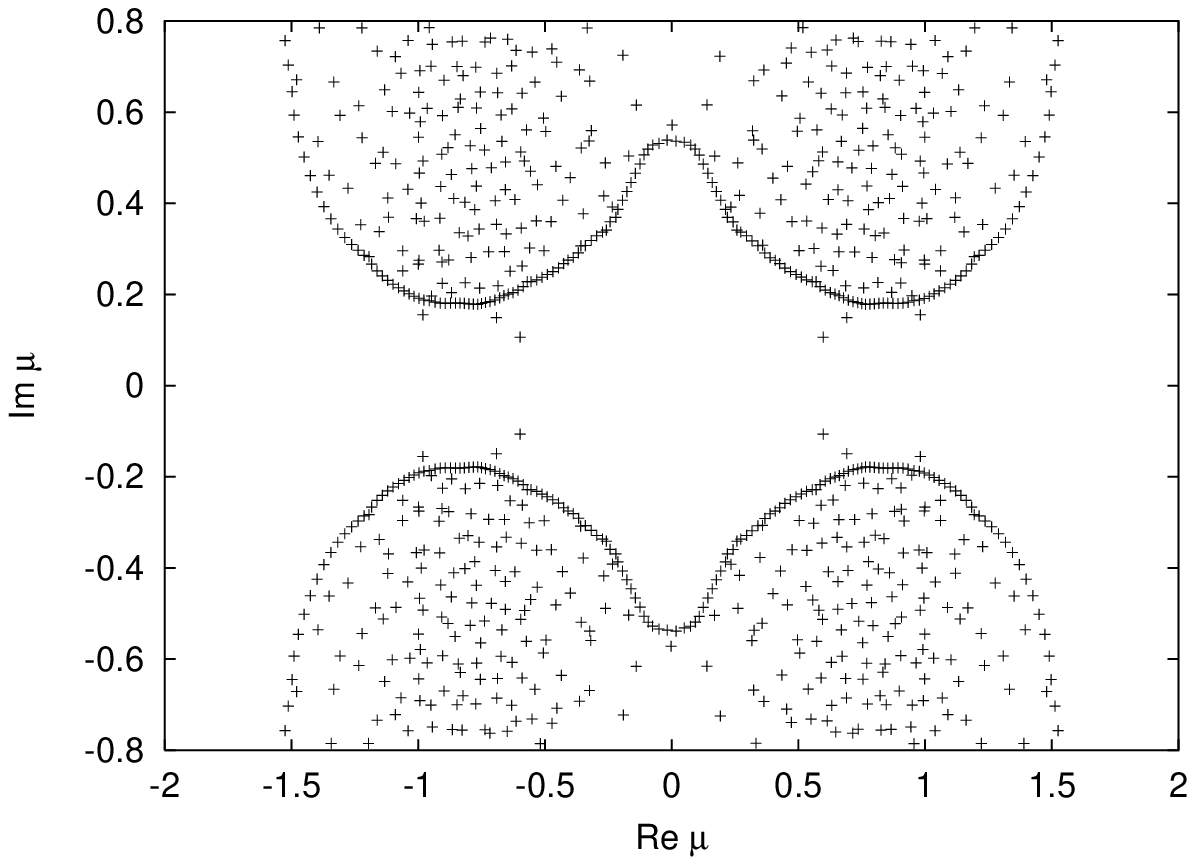, height=0.28\textheight}}
\vspace{-1.5cm}
\caption{Lee Yang zeros evaluated in the complex $\mu$ plane for a 
$6^{3}4$ lattice at $\beta=1.5$ from an ensemble generated at $\mu_{o}$=0.7.}
\end{figure}
\begin{table}
\begin{center}
$4^{4}\,\,\,$
\begin{tabular}{|c|c|c||} \hline
$\mu_{o} $ & Re $\eta_{1}$ & Im $\eta_{1}$ 
\\ \cline{1-1} \cline{2-2} \hline

0.3	& 0.502(0.109) & 0.117(0.171) \\ \hline

0.5	& 0.966(0.003) & 0.056(0.024) \\ \hline
	
0.7	& 0.871(0.066) & 0.098(0.103) \\ \hline

0.8	& 0.688(0.061) & 0.105(0.114) \\ \hline

0.9	& 0.824(0.072) & 0.237(0.077) \\ \hline

1.0	& 0.354(0.025) & 0.169(0.081) \\ \hline

1.1	& 0.560(0.015) & 0.142(0.069) \\ \hline

$\#. \, \rm{Ens.} $ & & 

\\ \cline{1-1} \cline{2-2} \hline

1	& 0.688(0.061) & 0.105(0.114) \\ \hline

3	& 0.556(0.002) & 0.015(0.025) \\ \hline

5	& 0.497(0.001) & 0.024(0.014) \\ \hline
	
7	& 0.480(0.001) & 0.014(0.013) \\ \hline

\end{tabular}	
\end{center}
\begin{center}
$6^{3}4$
\begin{tabular}{|c|c|c||} \hline
$\mu_{o} $ & Re $\eta_{1}$ & Im $\eta_{1}$ 
\\ \cline{1-1} \cline{2-2} \hline

0.3	 & 0.411(0.001) & 0.116(0.001) \\ \hline

0.5	 & 0.830(0.002) & 0.167(0.096) \\ \hline
	
0.7	 & 0.523(0.032) & 0.134(0.001) \\ \hline

0.8      & 0.822(0.028) & 0.154(0.082) \\ \hline

0.9	 & 0.546(0.067) & 0.153(0.051) \\ \hline

1.0	 & 0.434(0.039) & 0.091(0.039) \\ \hline

1.1	 & 0.461(0.011) & 0.064(0.030) \\ \hline

$\#. \, \rm{Ens.} $ & &

\\ \cline{1-1} \cline{2-2} \hline

1	& 0.546(0.067) & 0.153(0.051) \\ \hline

3	& 0.467(0.008) & 0.012(0.007) \\ \hline

5	& 0.453(0.008) & 0.011(0.007) \\ \hline
	
7	& 0.477(0.001) & 0.006(0.005) \\ \hline

\end{tabular}	
\caption{Lee Yang zero with the smallest imaginary part evaluated in the 
complex$\mu$ plane ($\eta_{n} = T \, ln \, \alpha_{n}$) for two lattice 
volumes at $\beta =1.5$. Dependence on value of $\mu_{o}$ used to 
generate ensembles for the Glasgow reweighting method (upper), and 
dependence on the number of ensembles included in the new composite 
reweighting 
scheme (lower).}
\end{center}
\end{table}
\begin{figure}
\centering{\epsfig{file=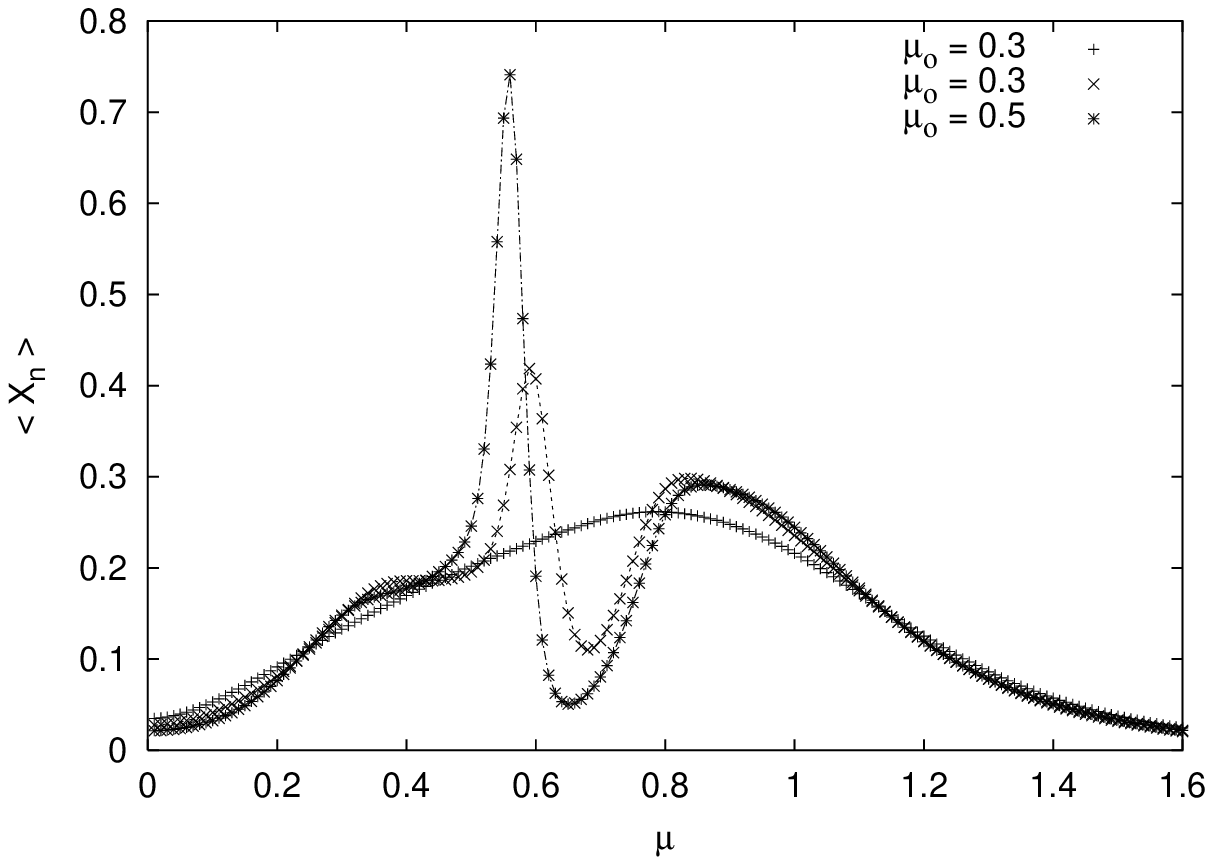, height=0.28\textheight}}
\centering{\epsfig{file=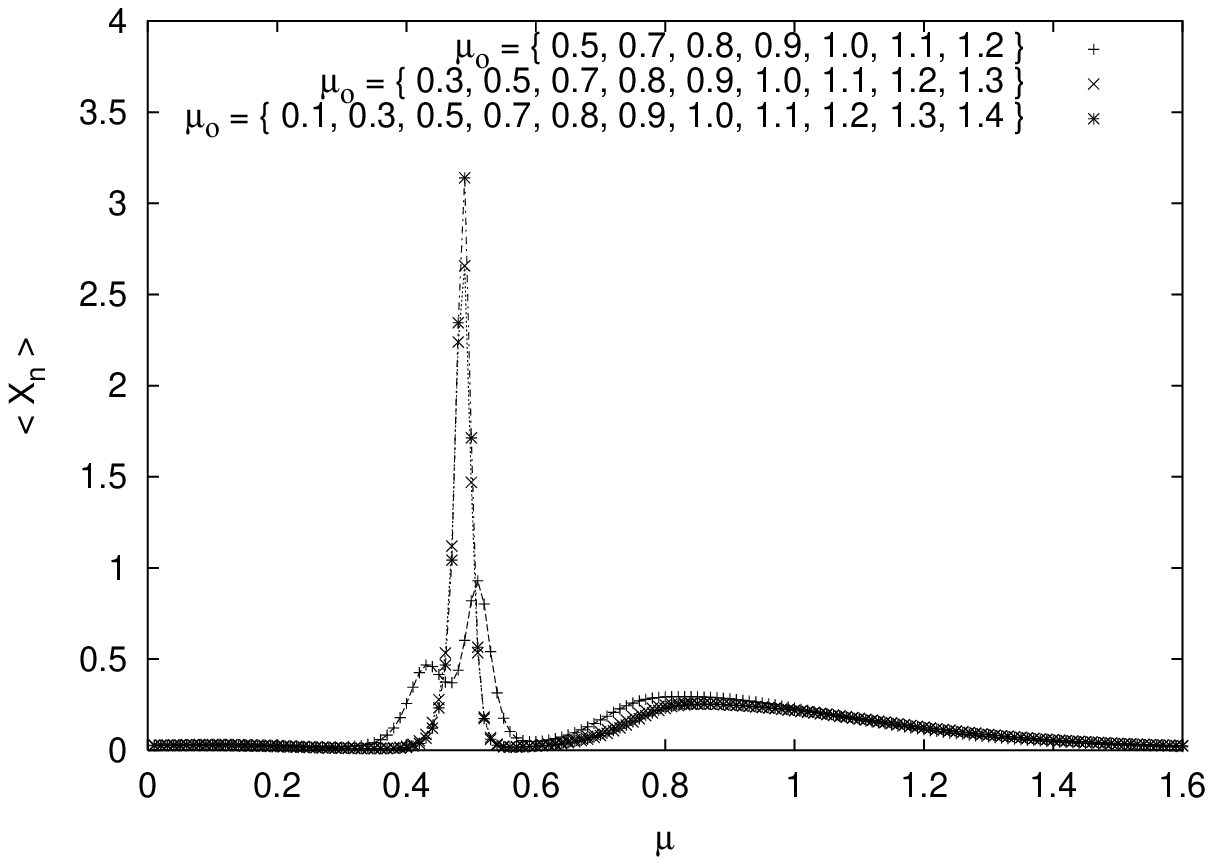, 
height=0.28\textheight}}
\caption{Quark number density susceptibility $\langle \chi_{n} (\mu) 
\rangle$
 for a $4^{4}$ lattice at $\beta = 1.5$, with an increasing number of 
ensembles 
included in the composite reweighting (upper 1, 3, 5) (lower 7, 9, 11). 
 A prominent peak develops as 
the number of composite reweighted splines is increased, indicating a first 
order transition.}
\end{figure}

\subsection{Weak Coupling}
There is a marked difference between the unphysical $\mu_{o}$ 
dependence of the Lee-Yang zeros associated with $\mu_{c}$ evaluated 
before composite reweighting at $\beta =1.5$ 
and $\beta = 2.3$ between Tables 2 and 3. In Table 2 there is some indication 
of competition between the two separate transition points during rootfinding 
we have identified 
above as the value of $\mu_{o}$ is varied, 
which is now entirely absent in Table 3. 
From this we can conclude that there is no indication of a 
second transition point at $\beta =2.3$, and that a transition can be readily 
identified at $\mu_{c} \sim 0.8$. Where the reweighting 
ensemble is generated at $\mu_{o} = \mu_{c}$ there is also good agreement 
between these zeros measurements and those from composite reweighting. 
Although again we have not quantified how small the difference between $\mu_{o}$ and $\mu_{c}$ must be for 
the reweighting to be effective, and have relied instead on composite 
reweighting.
   
Our composite reweighting Lee-Yang zeros and quark number density 
susceptibility measurements 
indicate again (where the zero closest the real $z$-axis 
goes to zero as the volume is increased and where 
$\langle \chi_{n} (\mu) \rangle$ becomes singular) that the 
transition is first order. The only context in which a first order 
transition is predicted in 
the effective chiral lagrangian approach 
is with a transition from the diquark to the symmetric phase for $\mu_{c} > 
\frac{1}{2}m_{\pi}$, and we therefore confirm the existence of such a 
transition line in the $T-\mu$ plane.

\begin{figure}
\centering{\epsfig{file=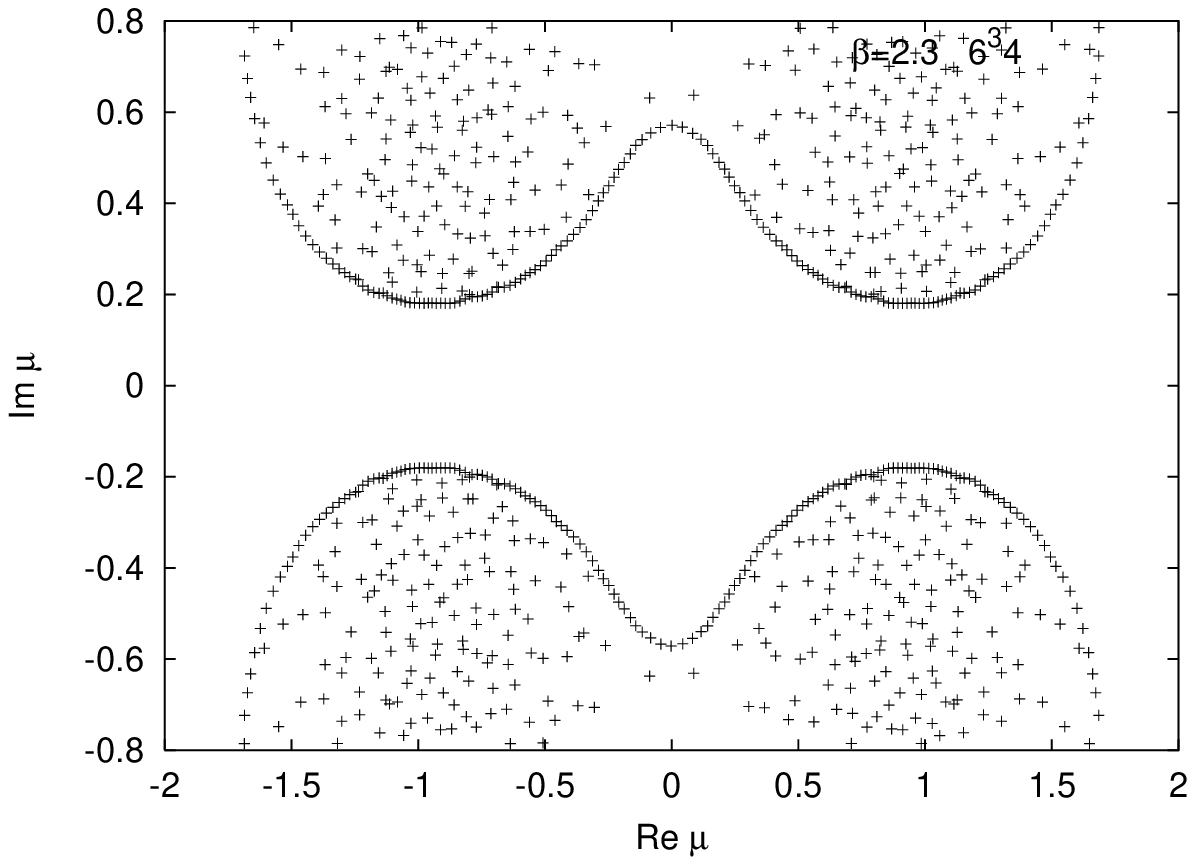, 
height=0.28\textheight}}
\vspace{-1.5cm}
\caption{Lee Yang zeros evaluated in the complex $\mu$ plane for a $6^{3}4$ 
lattice at $\beta=2.3$ from an ensemble generated at $\mu_{o}$=0.7.}
\centering{\epsfig{file=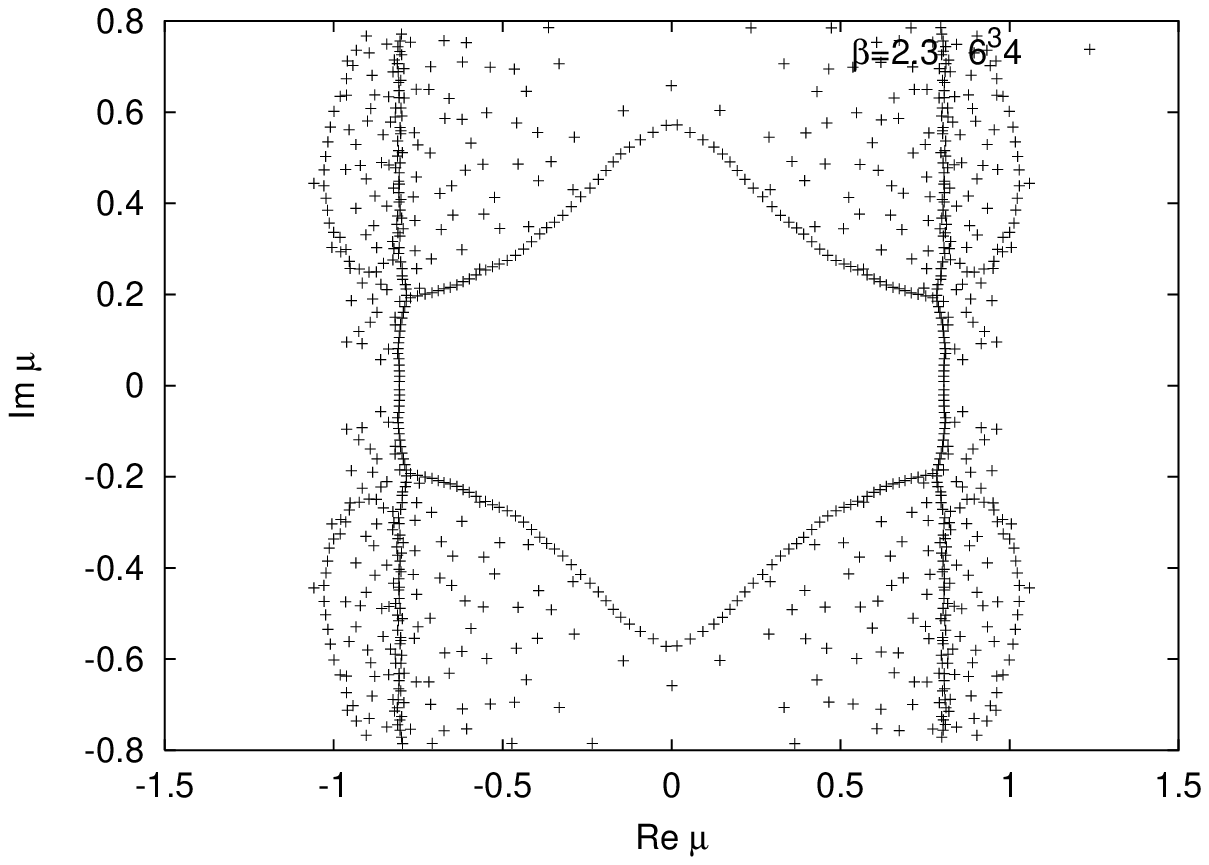, 
height=0.28\textheight}}
\vspace{-1.5cm}
\caption{Lee Yang zeros evaluated in the complex $\mu$ plane for a $6^{3}4$ lattice at $\beta=2.3$ from an ensemble generated at 
$\mu_{o}$=0.8.}
\centering{\epsfig{file=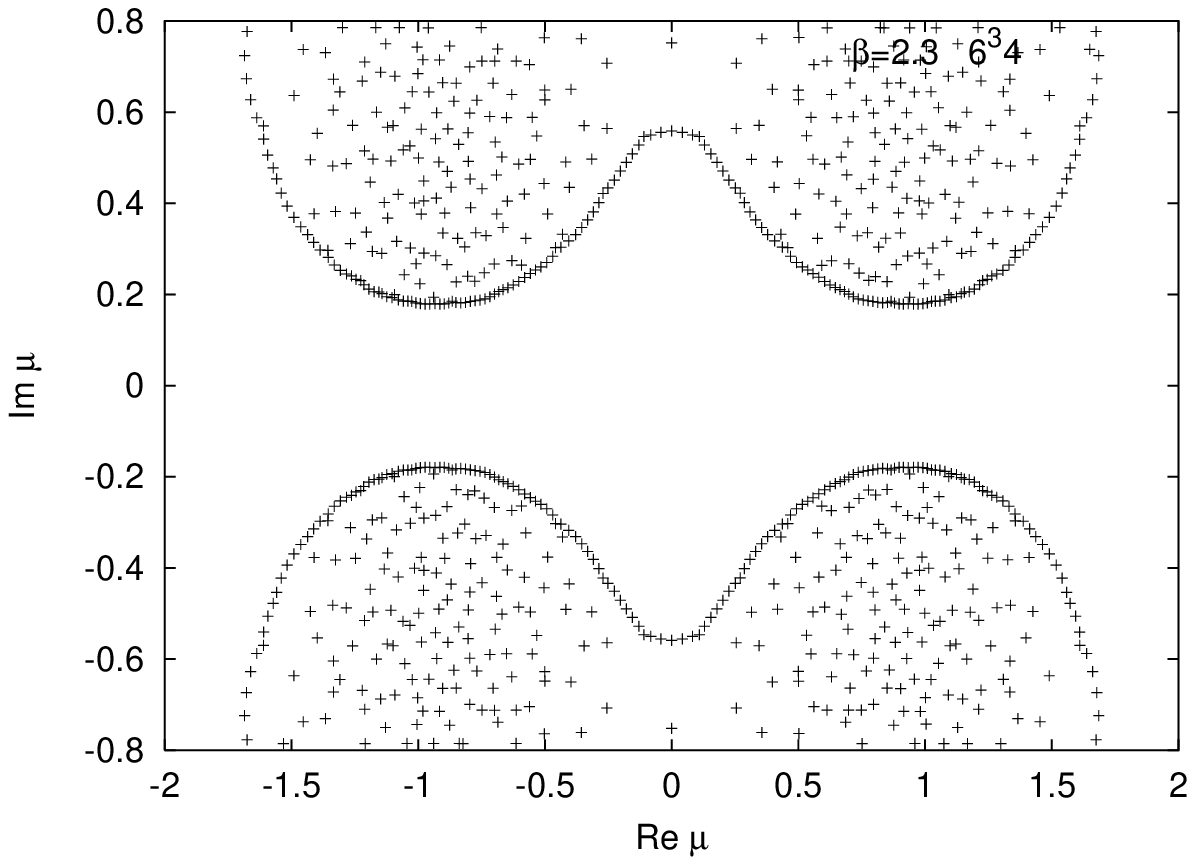, 
height=0.28\textheight}}
\vspace{-1.5cm}
\caption{Lee Yang zeros evaluated in the complex $\mu$ plane for a $6^{3}4$ lattice at $\beta=2.3$ from an ensemble generated at 
$\mu_{o}$=0.9.}
\end{figure}
\begin{table}
\begin{center}
$4^{4}\,\,\,$
\begin{tabular}{|c|c|c||}		\hline
$\mu_{o} $ & Re $\eta_{1}$ & Im $\eta_{1}$ 
\\ \cline{1-1} \cline{2-2} \hline
0.3	& 0.816(0.041) & 0.214(0.096) \\ \hline
0.5	& 0.801(0.059) & 0.235(0.059) \\ \hline
0.7	& 0.791(0.090) & 0.223(0.106) \\ \hline
0.8	& 0.797(0.042) & 0.099(0.135) \\ \hline
0.9	& 0.747(0.048) & 0.230(0.089) \\ \hline
1.0     & 0.734(0.040) & 0.200(0.094) \\ \hline
1.1	& 0.610(0.003) & 0.167(0.091) \\ \hline
$\# \, \rm{Ens.} $ & &    
\\ \cline{1-1} \cline{2-2} \hline
2	& 0.835(0.029) & 0.113(0.038) \\ \hline
4	& 0.839(0.005) & 0.082(0.036) \\ \hline
6	& 0.830(0.003) & 0.040(0.034) \\ \hline
8	& 0.849(0.005) & 0.031(0.026) \\ \hline
\end{tabular}
\end{center}
\begin{center}	
$6^{3}4$
\begin{tabular}{|c|c|c||}		\hline
$\mu_{o} $ & Re $\eta_{1}$ & Im $\eta_{1}$    
\\ \cline{1-1} \cline{2-2} \hline
0.3	& 1.144(0.028) & 0.144(0.060) \\ \hline
0.5	& 0.874(0.036) & 0.176(0.086) \\ \hline
0.7	& 0.883(0.039) & 0.176(0.093) \\ \hline
0.8	& 0.806(0.001) & 0.009(0.004) \\ \hline
0.9	& 0.907(0.021) & 0.319(0.033) \\ \hline
1.0     & 1.068(0.025) & 0.148(0.026) \\ \hline
1.1	& 0.935(0.002) & 0.172(0.014) \\ \hline
$\# \, \rm{Ens.} $ & &    
\\ \cline{1-1} \cline{2-2} \hline
2	& 0.807(0.001) & 0.009(0.008) \\ \hline
4	& 0.806(0.001) & 0.009(0.005) \\ \hline
6	& 0.803(0.001) & 0.007(0.004) \\ \hline
8	& 0.802(0.001) & 0.007(0.003) \\ \hline
\end{tabular}	
\caption{Lee Yang zero with the smallest imaginary part evaluated in the 
complex$\mu$ plane ($\eta_{n} = T \, ln \, \alpha_{n}$) for two lattice 
volumes at $\beta =2.3$. Dependence on value of $\mu_{o}$ used to 
generate ensembles for the Glasgow reweighting method (upper), and 
dependence on the number of ensembles included in the new composite 
reweighting 
scheme (lower).}
\end{center}
\end{table}
\section{Conclusions}
Despite expectations, generating an ensemble for the reweighting method with 
a value of $\mu_{o}$ arbitrarily close to $\mu_{c}$ still leads to an overlap 
problem in a model with a pseudoreal representation evaluated at $\beta_{c}$. 
In fact, for an exploratory study (in which $\mu_{c}$ is unknown), 
the fugacity expansion coefficients are more effectively sampled through the 
combination of terms from a covering series of ensembles.
Even with the real Monte Carlo measure of two colour QCD the overlap 
problem is still pathological for the Glasgow reweighting method. With 
new multi-parameter ($\beta$, $\mu$)
 reweighting approaches to SU(3) \cite{20} we 
would therefore expect there to be similar sampling bias in the coefficients 
across the full fugacity expansion.  
In order to successfully investigate the 
possibility of there being similar tricritical behavior in the $\mu - T$ 
phase plane of SU(3) \cite{21} with the Glasgow method, this 
issue of effective 
sampling across the full range of the polynomial this overlap problem 
should then be addressed. As we have, seen substantially different transition 
points are determined from reweighting measurements at intermediate coupling 
from ensembles generated with $\mu_{o}$ arbitrarily close to $\mu_{c}$.
 
Lee Yang zero analysis allows the simple identification of a first order 
transitions in $\mu$ with lattice measurements on comparatively small volumes. 
To extend the rigor of 
this approach and determine the critical exponents of the transition at 
$\mu_{c}$, however, it will be necessary to increase the lattice size 
as the zeros scaling at $\beta = 2.3$ is not without finite volume effects. 
It will also be interesting to repeat 
this volume scaling analysis for $\beta < \beta_{c}$ and to compare the 
measured critical exponents for the diquark phase transition with those
 predicted with the chiral langragian approach \cite{10}, 
believed to be second order.\\

Thanks to M. Alford for useful discussions.

\end{document}